\documentclass[journal=jacsat,manuscript=article]{achemso}

\usepackage[version=3]{mhchem} 
\usepackage{textcomp}
\usepackage{xcolor}

\author{Salvatore Costanzo}
\altaffiliation{Current address: DICMaPI, University of Naples, P.le Tecchio 80, Naples 80125, Italy}
\author{Amélie Banc}
\author{Ameur Louhichi}
\author{Edouard Chauveau}
\affiliation[L2C]
{Laboratoire Charles Coulomb (L2C), Univ. Montpellier, CNRS, Montpellier, France}
\author{Baohu Wu}
\affiliation[KWS-3]{Forschungszentrum Jülich GmbH JCNS am MLZ Lichtenbergstr. 1
85748 Garching}
\author{Marie-Hélène Morel}
\affiliation[INRA]
{Ing\'{e}nierie des Agro-polym\`{e}res et Technologies Emergentes (IATE), Univ. Montpellier, CIRAD, INRAE, Montpellier SupAgro, Montpellier, France}
\author{Laurence Ramos}
\affiliation[L2C]
{Laboratoire Charles Coulomb (L2C), Univ. Montpellier, CNRS, Montpellier, France}
\email{laurence.ramos@umontpellier.fr}
\phone{+33 0467144284}

\title[]
  {Tailoring the viscoelasticity of polymer gels of gluten proteins through solvent quality}

\abbreviations{IR,NMR,UV}

\begin{document}


\begin{abstract}

We investigate the linear viscoelasticity of polymer gels produced by the dispersion of gluten  proteins in water:ethanol binary mixtures with various ethanol contents, from pure water to $60$ \% v:v ethanol. We show that the complex viscoelasticity of the gels exhibits a time/solvent composition superposition principle, demonstrating the self-similarity of the gels produced in different binary solvents. All gels can be regarded as near critical gels with characteristic rheological parameters, elastic plateau and characteristic relaxation time,  which are related one to another, as a consequence of self-similarity, and span several orders of magnitude when changing the solvent composition. Thanks to calorimetry and neutron scattering experiments, we evidence a co-solvency effect with a better solvation of the complex polymer-like chains of the gluten proteins as the amount of ethanol increases. Overall the gel viscoelasticity can be accounted for by a unique characteristic length characterizing the crosslink density of the supramolecular network, which is solvent composition-dependent. On a molecular level, these findings could be interpreted as a transition of the supramolecular interactions, mainly H-bonds, from intra- to interchains, which would be  facilitated by the disruption of hydrophobic interactions by ethanol molecules. This work provides new insight for tailoring the gelation process of complex polymer gels.
\end{abstract}

\section{Introduction}

Solvation of macromolecules in water and in mixtures of solvents, and its link with the material properties (mechanical properties, in particular), is of central relevance for many areas of chemical physics, polymer physics, soft matter science and material science, but also in many applications, such as food and oil industry, polymer processing, pharmaceutics and cosmetics. In this framework, water:ethanol binary mixtures are solvents that should deserve a particular attention. On the one hand, from an industrial point of view, these binary mixtures are interesting  because of their low toxicity, biocompatibility, environment friendliness and food grade nature. On the other hand, from a more fundamental point of view, these mixtures are challenging solvents because they exhibit many abnormal properties due to complex molecular structuring, which originates from the differences in energy of hydrogen bonding between water-water, water-ethanol, and ethanol-ethanol molecules (see e.g. Refs.~\cite{dolenko_raman_2015,halder_unravelling_2018} and references therein). In general, binary solvents can modulate the aqueous solubility and the stability of polymers and proteins, through a modification of hydrophobic interactions and/or H-bonds. As a result, the solubility of polymers in binary solvent mixtures can result in poorer solubility  (co-nonsolvency) or, by contrast, in  improved solubility (co-solvency) compared to the individual solvents.  In water:ethanol binary mixtures, co-nonsolvency has been identified for common polymers as polyethylene oxide\cite{wang2014}, polyacrylamide and its derivatives~\cite{asadujjaman_polyacrylamide_2018}, in particular poly(N-isopropylacrylamide) PNIPAM~\cite{Costa2002, hore_co-nonsolvency_2013, bischofberger_co-nonsolvency_2014}, and also for bio-inspired polymers as elastin-like polypeptides~\cite{mills_cononsolvency_2019}. By contrast, water:ethanol binary mixtures can also facilitate the polymer solvation, as for instance for polymethyl metacrylate (PMMA),  for which it is believed that the disruption of the water clusters by ethanol molecules favors PMMA solvation~\cite{hoogenboom_solubility_2010}. Co-solvency rules also the solvation
of more complex polymeric species, comprising groups with different polarity and hydrophobicity. In these cases, co-solvency  can be viewed as a consequence of the polymer maintaining the optimum solvent environment, which will ensure maximum compatibility among the species (see for instance the review~\cite{zhang_polymers_2015} and references therein). These effects are also important for proteins, which are certainly among the most complex polymers one can imagine.  Here, water:ethanol mixtures impact the conformational states of the proteins because of the role of preferential solvation of protein hydrophobic residues by the ethyl groups of the ethanol molecules~\cite{ortore_preferential_2011,ghosh_solvent_2013,ghosh_composition_2015,amin_effect_2016,avdulov_direct_1996}. Overall, co-solvency in proteins and in polymers has been evidenced and investigated in many instances, but no general rules have been identified so far and the phenomenology appears to be very system specific, and thus very difficult to predict.

Although solvent quality, and co-solvency or co-nonsolvency, are related to the properties of individual polymer chains, they obviously have an effect on semi-dilute and concentrated polymer solutions and on polymer gels. The latter are formed by inducing chemical or physical crosslinking between polymer molecules in solution, leading to a process generally referred to as gelation.
Gelation of polymers is ubiquitous and has many important applications. Examples of chemical gels can be found in rubber industry, whereas many products in food, cosmetics, drug delivery and tissue engineering fields are based on physical gels. Regarding physical gelation, the solvent quality directly affects the strength and number density of supramolecular interactions on which the gel formation relies, for instance hydrogen bonding or hydrophobic interactions. Therefore, the solvent is expected to play a relevant role in determining gelation kinetics, structure and viscoelastic properties of the resulting gel.
In this perspective, investigating the link between polymer gelation and solvent quality is relevant as it may provide the tools to modulate the viscoelasticity of the gel by changing the nature of the solvent.
Some studies investigate how the solvent composition influences the overall viscoelasticity of polymer samples.  In Ref.~\cite{lai_influence_2013} chemical crosslinking of gelatin is induced by an external additive  whose solubility depends on the water:ethanol content, hence yielding gels whose viscoelasticity depends on the solvent composition. Quite similarly, it has been recently reported that the solvent composition tunes in a controlled fashion the strength of the hydrogen bonds-based crosslinks between two polyelectrolytes, leading to frequency shifts of the frequency dependent complex moduli of polymer complexes~\cite{mathis_tuning_2018}. In these examples, the solvent composition influences uniquely the strength and/or amount of crosslinks between polymer chains, presumably not the chain conformation itself. In this framework, studies on co-solvency or co-nonsolvency effect on polymer gels are rather scarce. Relevant works concern the swelling behavior of polymer gels and show that a re-entrant volume transition between swollen and collapsed states occurs when changing the solvent composition~\cite{katayama_reentrant_1984, amiya_reentrant_1987, hirotsu_critical_1988, ikkai_swelling_2003,boyko_preparation_2003}. This is a trivial phenomenon expected for gels with a fixed crosslink density between polymer chains, whose conformation varies from good solvents to bad solvents. In this simple situation, we expect the elastic modulus of the gel to vary in direct relation to the crosslink density, although, to the best of our knowledge, rheological properties have not been probed yet.
Water:ethanol binary mixtures have also been used to modulate the sol-gel transition of bio-polymeric systems. For instance, it has been reported that the increase of ethanol content promotes the gelation of glycylalanylglycine in water:ethanol binary  solvents~\cite{thursch}. In this case, ethanol induces hydrophobic interactions between peptides by decreasing the solubility in water. A similar mechanism has been proposed for the gelation of a globular protein, $\beta$-lactoglobulin, in water:ethanol binary mixtures~\cite{dufour1998}, where protein denaturation in the presence of ethanol molecules induces gelation. But in these cases, the analogy of the resulting gels with polymer gels is questionable due to the fibrillar nature of the networks. In conclusion the links between solvent quality and rheological properties of polymer gels have not been clearly established so far, even for the simplest polymer gels. In this framework, we study here how solvent quality affects the viscoelasticity of near critical gels, a generic class of polymer gels.

We investigate gels that are produced by dispersing gluten proteins in binary water:ethanol mixtures with different compositions, ranging from pure water to $60$ \% v:v ethanol. Gluten comprises a $50/50$ w:w mixtures of monomeric proteins (gliadins) and long flexible branched  polymer-like proteins (glutenins) resulting from glutenin sub-units covalently linked together by disulfide bonds\cite{LINDSAY1999,wieser_chemistry_2007}. These proteins possess intrinsically disordered domains conferring them flexibility. Using scattering techniques, we have previously shown that gluten proteins dispersed in a water:ethanol mixture with $50$ \%  v:v ethanol behave as flexible branched polymer chains (persistence length of the order of $0.7$ nm) in good solvent, both in the dilute and semi-dilute regimes~\cite{dahesh2014}. In addition, we have evidenced, still for a $50$ \%  v:v ethanol solvent,  an aging time- and concentration-dependent sol-gel transition, due to the building up of a polymer network held by H-bonds, and quantitatively rationalized the phenomenon in the framework of the near critical gel model developed initially for synthetic polymers~\cite{dahesh_spontaneous_2016}. Hence gluten gels display most of the structural and rheological features of polymer gels. In the present work, we show that, at a fixed gluten concentration, the solvent composition plays a key role in determining the gluten gel viscoelasticity. Akin to time/concentration superposition, we develop a time/solvent composition superposition principle, which allows one to correlate the gelation of gluten to the solvent quality for the polymer-like proteins. We also provide  a possible molecular understanding based on the balance between intra- and intermolecular H-bonds, which is tuned by solvent composition, through the disruption of hydrophobic interactions between gluten chains by ethanol molecules.

The paper is organized as follows. We first describe the materials and methods. We then present our experimental results on gel viscoelasticity. These measurements are complemented by calorimetry and scattering measurements, in order to get physical insights on the role of solvent composition on the gel structure and viscoelasticity, which are discussed in the last section of the manuscript.

\section{Materials and Methods}
\subsection{Materials}
\subsubsection{Gluten fractionation}
Native gluten powder is kindly supplied by Tereos Syral (France). Model gluten is obtained from native gluten based on a fractionation protocol~\cite{JustineThesis} adapted from Ref.~\cite{dahesh2014}. In brief, native gluten is dispersed in a water:ethanol binary mixture with $50$ \%  v:v ethanol. The dispersion is stirred for $19$ h at $20$\textdegree C and then centrifuged ($30$ min, $15000$ g). The supernatant is recovered and kept at $2$\textdegree C for $1$ h, yielding a liquid-liquid phase separation into two phases. We collect the dense phase, freeze it at $-40$\textdegree C, and then freeze-dry and grind it. The resulting powder constitutes the model gluten used in this work. Chromatography in a denaturating solvent indicates that model gluten is composed of an equal balance in mass of the two main classes of gluten proteins, monomeric gliadins (molecular weight in the range $11-90$ kg/mol) and polymeric glutenins (molecular weight in the range $90-6000$ kg/mol) ~\cite{dahesh2014, morel_insight_2020}.

\subsubsection{Sample preparation}
Water:ethanol mixtures are prepared by weighing proper amounts of water and ethanol, and mixing them in a plastic vial at ambient conditions. In order to achieve a specific volume fraction of ethanol, we consider a water density of $997.1$ g/cm$^3$ and an ethanol density $0.786$ g/cm$^3$. Solvent mixtures with ethanol content of $0$, $10$, $20$, $30$, $40$, $50$, and $60$ \% v:v are used. The density of the mixtures varies from $0.997$ to $0.889$ g/cm$^3$, as calculated according to reference~\cite{khattab2012}. To prevent bacterial growth, sodium azide ($1$ g/L) is added to the sample prepared with pure water.

\begin{figure}[htbp]
\centering
\includegraphics[width=1\textwidth]{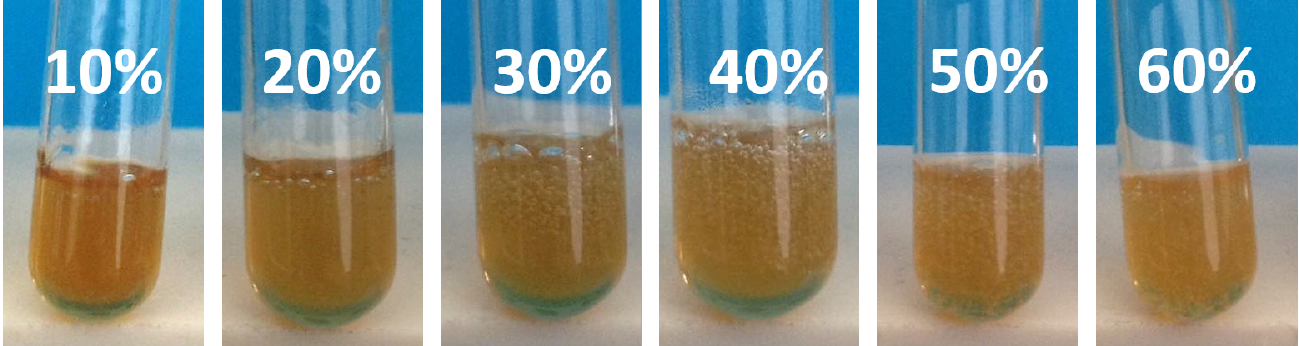}
\caption{\label{fig1} Images of gels  with different ethanol contents, as indicated in the legend. Images are taken at room temperature, $30$ days after sample preparation.}
\end{figure}

For each solvent composition, three different samples with a fixed gluten concentration of $0.5$ g/cm$^3$ are prepared according to the following protocol: approximately $0.5$ g of model gluten powder are inserted in a glass vial, to which the required amount of solvent is added to reach a protein concentration of $0.5$ g/cm$^3$. The mixture is gently stirred with a spatula for approximately one minute, in order to favor a proper mixing between the protein powder and the solvent. The vial is then closed, sealed with parafilm and put on a rotor for $24$ h to achieve homogenization. The vial is stored at $25$\textdegree C. After approximately $20$ days, all samples look like homogeneous and translucent gels with a yellowish color (Figure~\ref{fig1}). Note that, for solvent comprising more than $60$ \% ethanol v:v, the samples are very heterogeneous, presumably because of protein denaturation, thus preventing any reliable measurements.  Previous experiments  with a $50:50$  v:v water:ethanol binary mixture~\cite{dahesh_spontaneous_2016} have shown a concentration-dependent gelation kinetics with a characteristic time to reach an equilibrated steady gel of about $4$ days, for a concentration of $0.37$ g/cm$^3$, hence slightly smaller than the samples investigated here. This characteristic time has been found to decrease when concentration increases. Accordingly, and to ensure that equilibration has been attained even with solvents with low ethanol contents, we have chosen to start measurements after $30$ days of aging.

For very small angle neutron scattering experiments, gluten samples are prepared at a concentration of $0.462 \rm{g/cm}^3$. Hydrogenated samples are prepared with a solvent composed of milliQ water with $0.1$ \% sodium azide and ethanol (C$_2$H$_5$OH, analytical grade), and deuterated ones with a solvent composed of D$_2$0 (isotopic purity $> 99.97$ \%) with $0.1$ \% sodium azide and partially deuterated ethanol (C$_2$H$_5$OD, isotopic purity $> 99.97$ \%).


\subsection{Experimental techniques}
\subsubsection{Rheological measurements}
Rheological measurements are performed on a stress-controlled MCR501 rheometer  (Anton Paar, Germany). Standard stainless-steel parallel plates with a diameter of 25 mm are used as measuring geometry. Temperature control is ensured by means of a Peltier element (PTD-200). The measuring temperature is $25$\textdegree C. Prior to each measurement, the sample is loaded and left at rest for $20$ min, in order to ensure sample relaxation and thermal equilibration. The rim of the sample is covered with a low viscosity silicon oil ($0.1$ Pas) to minimize evaporation. Strain sweep tests are carried out before evaluating the frequency response, in order to assess the linear viscoelastic regime. All frequency sweeps are performed at a strain amplitude of $1$ \%, well in the linear viscoelastic regime.

\subsubsection{Differential scanning calorimetry}
Modulated temperature differential scanning calorimetry measurements are performed using the DSC Q2000 calorimeter (TA instruments, USA) calibrated with indium. For each protein solution, aluminum pans ($40$ $\mu$L TzeroHermetic Pan and Lid) are prepared with $10$-$30$ mg of sample and sealed. An empty pan is used as reference. The sample is first equilibrated at $-20$ \textdegree C for $5$ min. A modulated temperature with an amplitude of  $0.3$ \textdegree C over a period of $60$ s is then applied with a heating rate of $2$ \textdegree C/min from $-20$ \textdegree C to $40$ \textdegree C. At the end of the heating step, the sample is maintained for $10$ min at $40$ \textdegree C, before the same modulated temperature (amplitude of  $0.3$ \textdegree C over a period of $60$ s) is applied with a cooling rate of $-2$ \textdegree C/min down to $-20$ \textdegree C. Reported values (onset temperature of the transition and amplitude of the thermal transition) are obtained with the TA Universal Instrument Analysis software (version 4.5A).

\subsubsection{Very small angle neutron scattering measurements}
Very small angle neutron scattering (VSANS) experiments have been performed on the KWS3 instrument running on the focusing mirror principle~\cite{heinz2015,radulescu2016} operated by the J\"ulich Center for Neutron Science at the Heinz Maier-Leibnitz Zentrum (MLZ, Garching Germany). A sample-to-detector distance of $1.95$ m is used with a
wavelength $\lambda$= 12.8 \AA, giving access to $q$-vectors from 10$^{-3}$ to 10$^{-2}$ \AA$^{-1}$. The samples are held in $1$ mm-thick quartz cells maintained at $30$\textdegree C. At this temperature, all (hydrogenated and deuterated) samples are in their monophasic domain.
Standard reduction of raw data is performed by the routine qtiKWS \cite{qtikws} including corrections for detector sensitivity, background noise and empty cell signal.

\section{Results and discussion}
\subsection{Experimental results}
\subsubsection{Linear viscoelasticity}

\begin{figure}[htbp]
\centering
\includegraphics[width=0.8\textwidth]{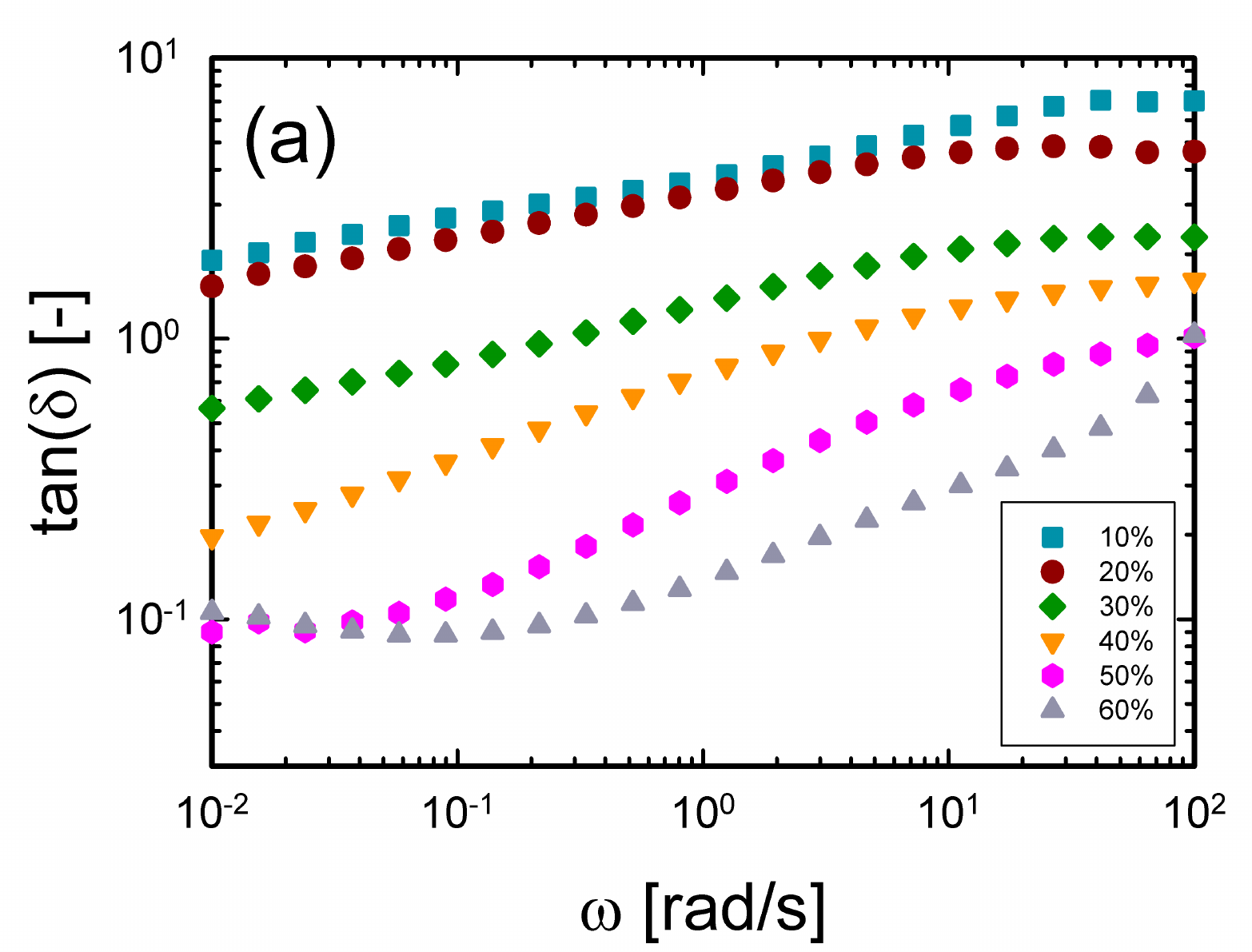}
\includegraphics[width=0.8\textwidth]{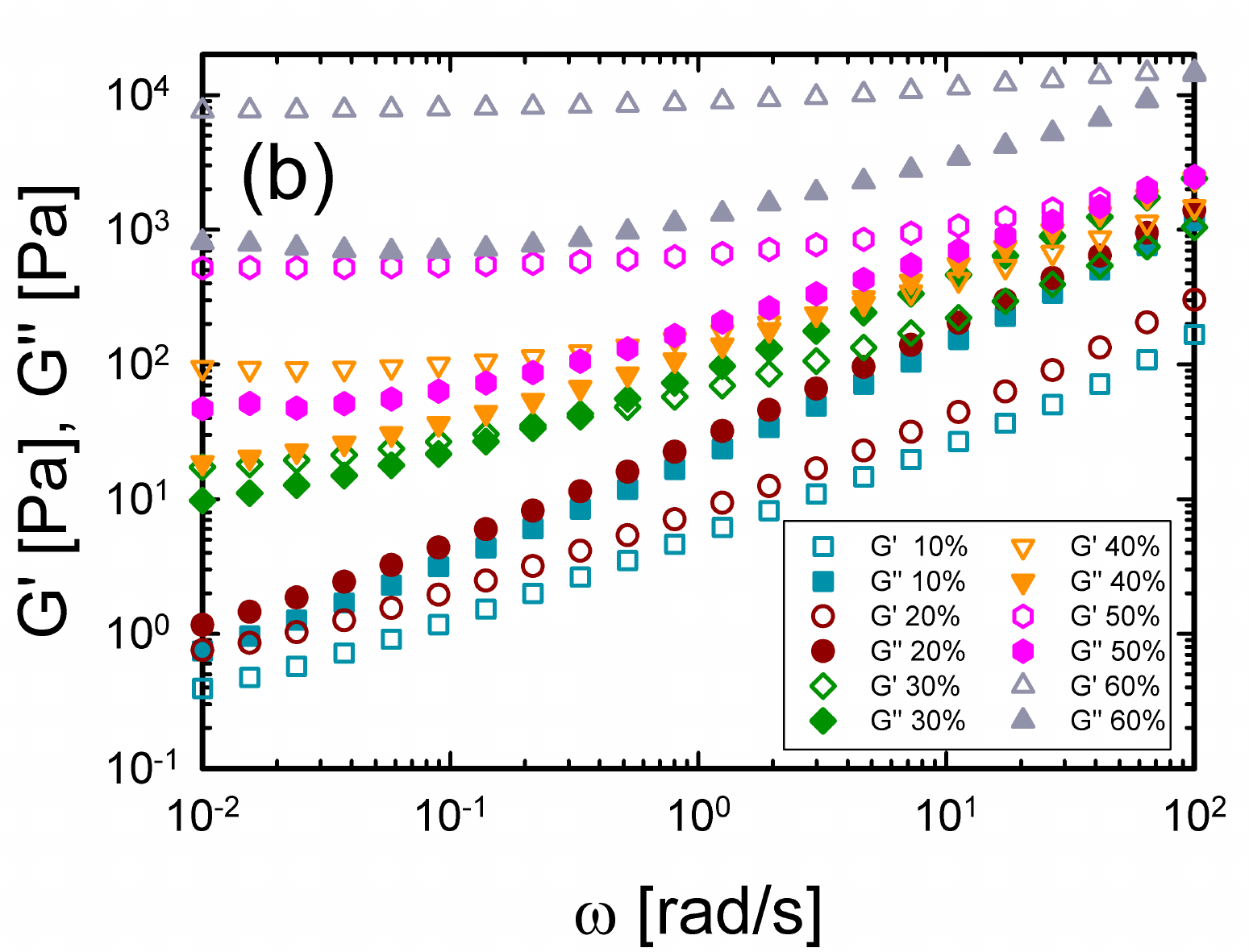}
\caption{\label{fig2} Frequency-dependent, (a) loss factor, $G"/G'$, and (b) storage, $G'$ (empty symbols), and loss, $G"$ (filled symbols), moduli,  for samples prepared with solvents having different ethanol contents as indicated in the legend. The strain amplitude is fixed at $1$ \% and the temperature at $25$\textdegree C. The gluten concentration is $0.5$ g/cm$^3$ and the aging time is $30$ days.}
\end{figure}

Figure~\ref{fig2} shows the dynamic frequency sweep tests on model gluten dispersed in different solvents, $30$ days after sample preparation. The different samples are already in the post-gel regime as proved by the fact that the loss factor, $\tan \delta=\frac{G"}{G'}$, with $G'$ the storage modulus, and $G"$, the loss modulus, increases with frequency for all of them~\cite{Joshi2020} (Figure~\ref{fig2}(a)). A slight upturn is observed at low frequency only for the sample with $60$ \% v:v ethanol, suggesting that flow will be eventually approached at extremely long time-scales. The distance from the critical gel point is not the same for all samples, as proved by their different viscoelasticity. Markedly different behaviors are measured for the storage, $G'$, and the loss, $G"$, moduli, depending on the solvent composition used (Figure \ref{fig2}(b)). Samples comprising small amounts of ethanol ($10$ and $20$ \% v:v) are much closer to the critical gel state where $G'\propto G''\propto \omega^{\Delta}$. For such samples, the loss and storage moduli are nearly parallel in a log-log plot in the explored frequency range. In particular, at high frequency, $G'$ and $G"$ exhibit a power-law dependence upon frequency with an exponent  $\Delta=0.83 \pm 0.01$ whereas, at low frequency, there is a slight tendency of $G'$ to form a plateau, indicating the formation of the incipient gel.
By contrast, the samples prepared with a solvent rich in ethanol ($40$, $50$ and $60$ \% v:v) display a well developed gel-like behavior with $G'$ larger than $G"$, and a low frequency elastic plateau, $G_0$, which strongly increases with the amount of ethanol in the solvent  (from $91$ Pa for a solvent with $40$ \% to $7700$ Pa for $60$ \% ethanol v:v).  Consistently, the sample prepared with a solvent with $30$ \% v:v ethanol displays an intermediate behavior.

\begin{figure}[htbp]
\centering
\includegraphics[width=0.85\textwidth]{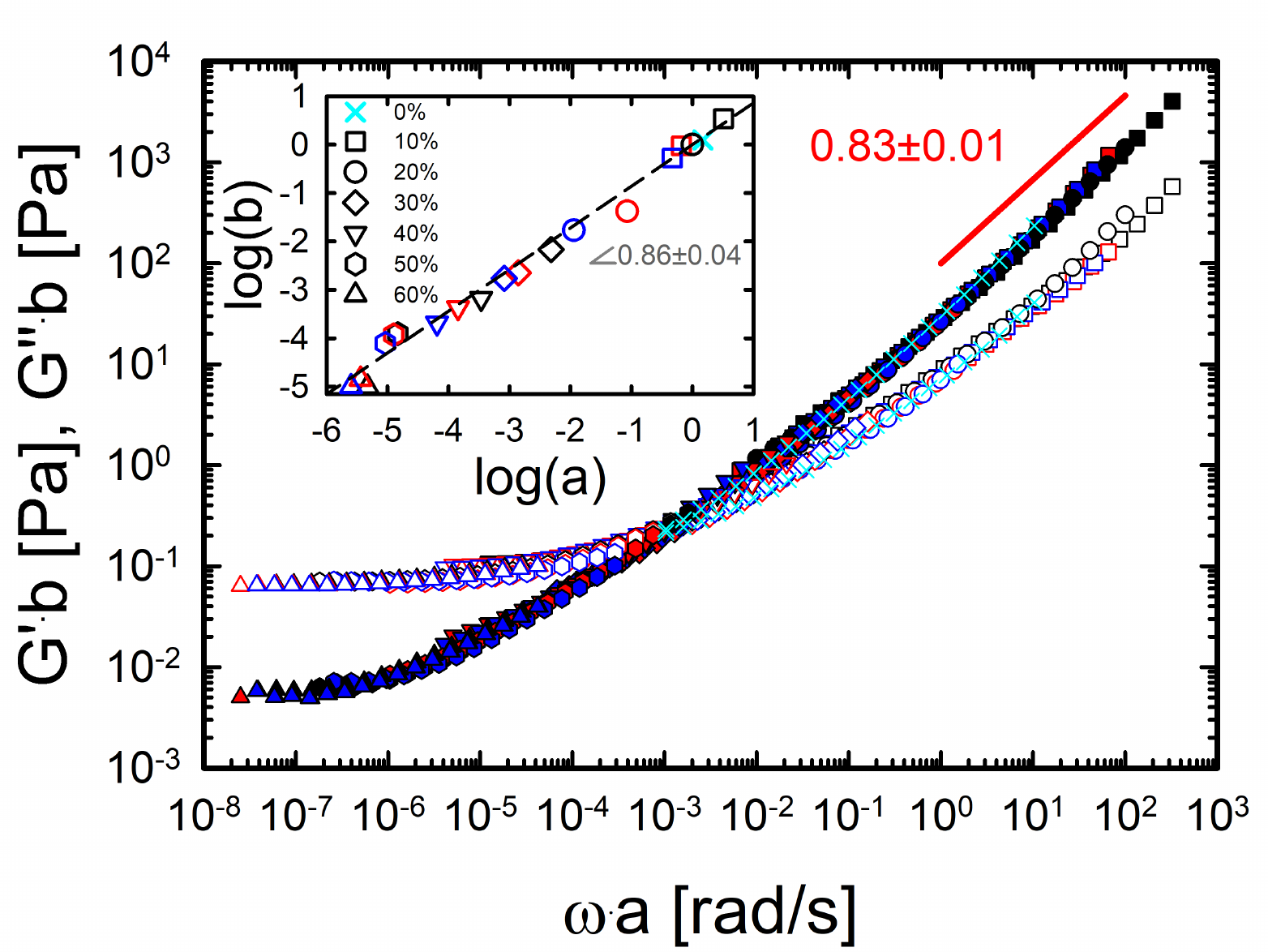}
\caption{\label{fig3} Re-scaled storage (empty symbols) and loss (filled symbols) moduli as a function of re-scaled frequency for all data taken at different aging times and different solvent compositions. Different symbols correspond to different ethanol content in the solvent: crosses, $0$\% v:v; squares, $10$\% v:v; circles $20$\% v:v; diamonds, $30$\% v:v; down triangles, $40$\% v:v; hexagons, $50$\% v:v; up triangles, $60$\% v:v. Different colors correspond to different aging times: black, $30$ days; cyan, $40$ days; red, $60$ days; blue, $90$ days. The frequency response of the solution with $20$ \% v:v ethanol after $30$ days of aging is used as reference. The inset reports the vertical shift factor, $b$, versus the horizontal shift factor, $a$. The plot is parametric in ethanol concentration and ageing time (we use the same symbols and color code as for the main figure.)}
\end{figure}

In a previous work, some of the authors observed that gluten gels in a water:ethanol solvent comprising $50$ \% v:v ethanol obey a time-concentration superposition principle\cite{dahesh_spontaneous_2016}, suggesting self-similarity of the gel network obtained at different gluten concentrations for a given solvent composition. Here, we extend such a principle, and demonstrate that the self-similarity of the gluten network is dictated by the solubility of gluten proteins in water:ethanol solvents with different composition.
The data in Figure~\ref{fig2}(b) are such that the viscoelastic curves of the different samples can be horizontally and vertically shifted to obtain master curves. In order to build robust and reliable master curves, the following protocol is applied. Experimental data for the loss factor
are first horizontally shifted to obtain the best superposition of the different experimental curves. Note that, if self-similarity holds for the viscoelastic data, we expect that both $G'$ and $G"$ vary in a self-similar way for the different samples. Hence, the loss factor, which is the ratio between these two quantities, is not prone to vertical shift. Once the horizontal shift factors are obtained from the shift of the loss factor, we apply them independently to the two viscoelastic moduli, $G'$ and $G"$, and then shift $G'$ and $G"$ vertically to obtain the best superposition. Figure~\ref{fig3} reports the master curves obtained with the data of Figure~\ref{fig2}, using as a reference the experimental data acquired with a solvent comprising $20$ \% v:v ethanol after 30 days of aging. Data at 60 and 90 days are also included. The plot of the loss and storage moduli as functions of $\omega a$, with $\omega$ the actual frequency and $a$, the shift factor, allows to span almost $11$ orders of magnitude for the viscoelastic behaviour. In Figure~\ref{fig3}, a sample prepared with pure water (aging time $40$ days) is also reported (cyan crosses). Interestingly, the data fall into the same master curve as for the samples at different water:ethanol composition, indicating that very weak gels can be also obtained in pure water, a solvent commonly accepted as being a bad solvent for gluten proteins.

The resulting master curves in Figure \ref{fig3} display distinctive hallmarks. The viscoelastic response is essentially elastic at low frequency, $G'>G"$, with the emergence of an elastic plateau about one order of magnitude larger than the loss modulus. At high frequency, by contrast, the viscoelasticity is dominated by the loss modulus and the two moduli vary as a same power law with frequency  $G' \sim G" \sim \omega^{\Delta}$, with a critical exponent $\Delta = 0.83\pm0.01$. Of note, the master curve is very similar to those theoretically expected and experimentally observed for near critical gels produced with synthetic polymers and biopolymers above their gelation point,  i.e. in the so-called post-gel regime  ~\cite{martin_viscoelasticity_1988,martin_viscoelasticity_1989}, as well as for solid networks made of weakly attractive colloidal particles\cite{Trappe2000}. Near critical gels, with chemical or physical bonds (see the review~\cite{winter_rheology_1997} and references therein), extend the pioneer work of Winter and Chambon beyond the gel point~\cite{WinterChambon1987}. At the gel point, the gel is said critical and characterized by a network that is self-similar at all length scales. For near critical gels, the characteristic length $\ell_{c}$, below which the simple picture of critical gel holds, is related to $\omega_c$ (defined as the frequency at which $G'$ and $G''$ cross) as $\ell_{c}\propto\omega_c^{-1/\Delta}$~\cite{vilgis1988}. In this regime the self-similarity decreases with the distance from the critical gel point. In the post-gel regime, the strength and dynamics of the network are dictated by the density and life-time of the supramolecular junctions. An increase of the modulus indicates an increase of the number density of intermolecular junctions in the system.  The characteristic crossover frequency $\omega_{c}$ increases consistently with increasing number density of junctions, according to the self-similarity of the network. For near critical gels, because of the fractal self-similar structure of the stress-bearing network, both the elastic plateau, $G_0$, and the characteristic cross-over frequency, $\omega_c$, are predicted to follow critical variations with the distance $\epsilon$ from the gel point~\cite{martin_viscoelasticity_1988,martin_viscoelasticity_1989}:  $G_0 \sim \epsilon^z$ and $\omega_c \sim \epsilon^y$, with critical exponents $z$ and $y$ such that $\Delta = z/y$. This leads to a power law relation between the two viscoelastic parameters, $G_0$ and $\omega_c$:
$G_0 \sim \omega_c^{z/y} = \omega_c^{\Delta}$. The critical exponent $\Delta$ is related to the fractal dimension of the stress-bearing network~\cite{muthukumar_screening_1989}.

\begin{figure}[htbp]
\centering
\includegraphics[width=0.8\textwidth]{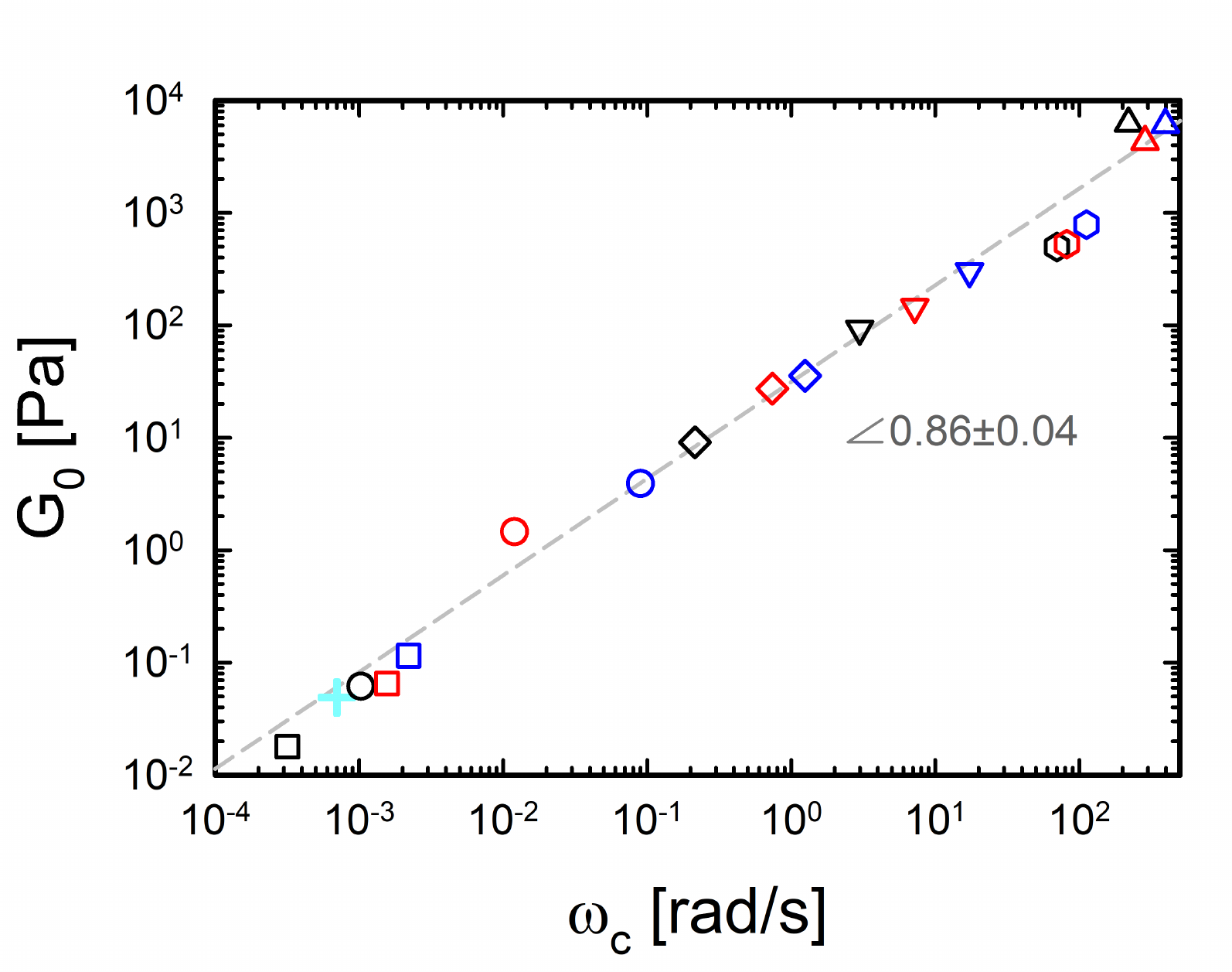}
\caption{\label{fig4} Elastic modulus as a function of characteristic frequency, as derived from the shift factors used to build the master curves shown in Figure~\ref{fig3}.  Different symbols correspond to different ethanol content in the solvent: crosses, $0$\% v:v; squares, $10$\% v:v; circles $20$\% v:v; diamonds, $30$\% v:v; down triangles, $40$\% v:v; hexagons, $50$\% v:v; up triangles, $60$\% v:v. Different colors correspond to different aging times: black, $30$ days; cyan, $40$ days; red, $60$ days; blue, $90$ days.}
\end{figure}

The inset of Figure~\ref{fig3} reports the vertical shift factor, $b$, as a function of the horizontal shift factor, $a$.
From the shift factors, $a$ for frequency and $b$ for the moduli, used to build the master curves, one can extract the elastic plateau, $G_0$, and the cross-over frequency, $\omega_c$, for all samples and aging times, even in the case where the rheological parameters cannot directly be measured experimentally, due to the limited frequency window experimentally accessible. By definition of the shift factors $\omega_c = \omega_c^{\rm{ref}} / a$ and $G_0=G_0^{\rm{ref}} / b$, with $\omega_c^{\rm{ref}}$  and $G_0^{\rm{ref}}$ being the numerical values for the reference sample ($20$\% v:v ethanol, $30$ days aging), as determined experimentally:  $\omega_c^{\rm{ref}}=1.03\times 10^{-3}$ rad/s and $G_0^{\rm{ref}}=6.2\times 10^{-2}$ Pa.
For all the different water-ethanol compositions, we plot the elastic plateau as a function of the critical frequency in Figure~\ref{fig4}. We find that all data fall onto a unique curve, with $G_0$ varying as a power law with $\omega_c$ with an exponent $0.86\pm0.04$. Note that, by definition, $b$ varies as a power law with $a$, with the same power law exponent as the $G_0$ versus $\omega_c$  plot (inset of Figure~\ref{fig3}). Remarkably, the exponent is equal within experimental error to the one found for the scaling of $G'$ and $G"$ at high frequency ($0.83\pm0.01$), in  agreement with the theoretical predictions for the viscoelasticity of near-critical gels.

\begin{figure}[htbp]
\centering
\includegraphics[width=0.8\textwidth]{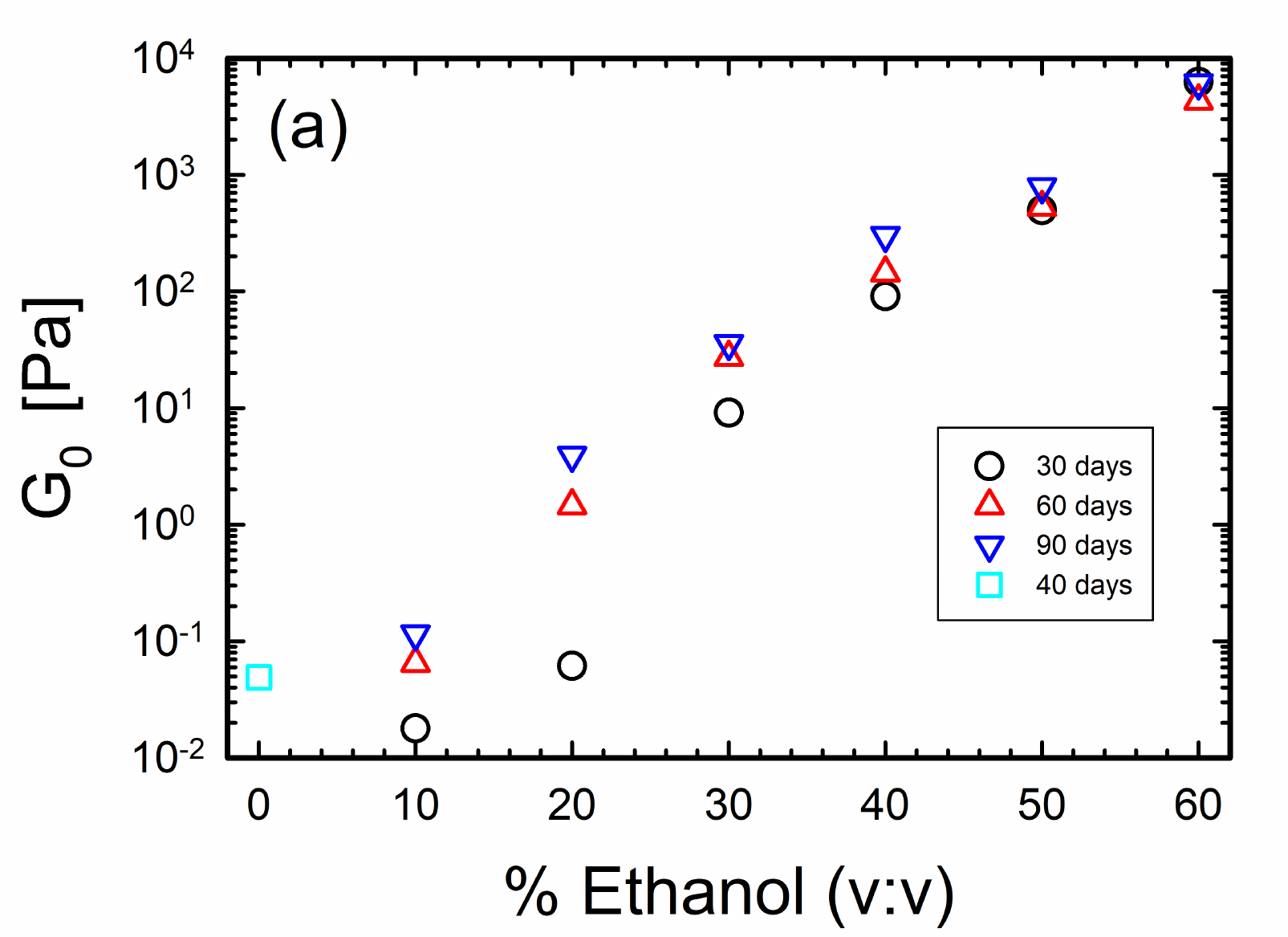}
\includegraphics[width=0.8\textwidth]{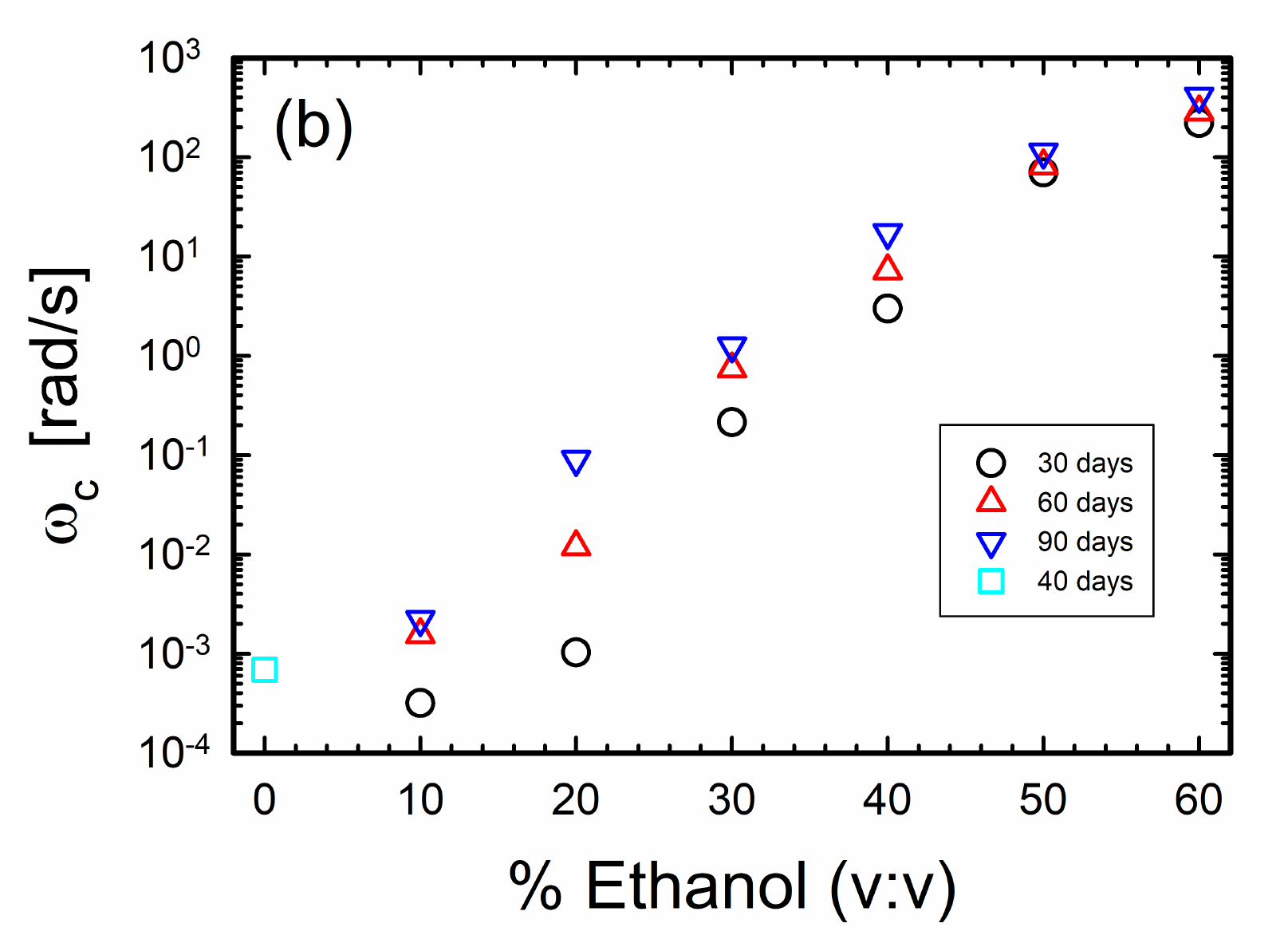}
\caption{\label{fig5} (a) Low frequency elastic plateau and (b) cross-over frequency as functions of ethanol content in the solvent for several sample ages, as indicated in the legend.}
\end{figure}

Figures~\ref{fig3} and~\ref{fig4}  demonstrate the applicability of a time-solvent composition superposition in the framework of near-critical gels, implying that the networks of gluten protein are self-similar with solvent composition. In addition, our results emphasize  the crucial role of the solvent composition in determining the viscoelasticity of gluten gels. This is further illustrated in Figure~\ref{fig5}(a) where the low frequency elastic plateau, $G_0$, is plotted as as function of the percentage of ethanol in the solvent used to prepare the samples. We find that $G_0$ increases continuously over $6$ orders of magnitude, from ca. $10^{-2}$ to $10^4$ Pa when the solvent is continuously changed from pure water to $60$ \% v:v ethanol. Similarly, $\omega_c$ increases from ca. $10^{-4}$ to $10^2$ rad/s  when the solvent is continuously changed from pure water to $60$ \% v:v ethanol (Figure~\ref{fig5}(b)). In the framework of near critical gels, the increase of $G_0$ and $\omega_c$ with the ethanol content of the solvent indicates that, as the percentage of ethanol in water/ethanol blend increases, the self-similar gluten networks move further and further away from their gel point. We note moreover an aging effect, with an elastic modulus and a cross-over frequency that increase with sample age, in accordance with our previous results~\cite{dahesh_spontaneous_2016}, but with a growth much more moderate than the one measured with solvent composition for most of the samples. Therefore, aging effects will not be discussed further in this work.

\subsubsection{Differential scanning calorimetry}
\begin{figure}[bp]
\centering
\includegraphics[width=0.72\textwidth]{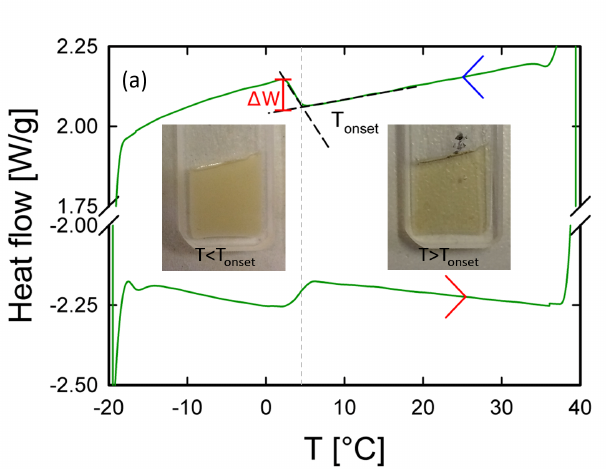}
\includegraphics[width=0.7\textwidth]{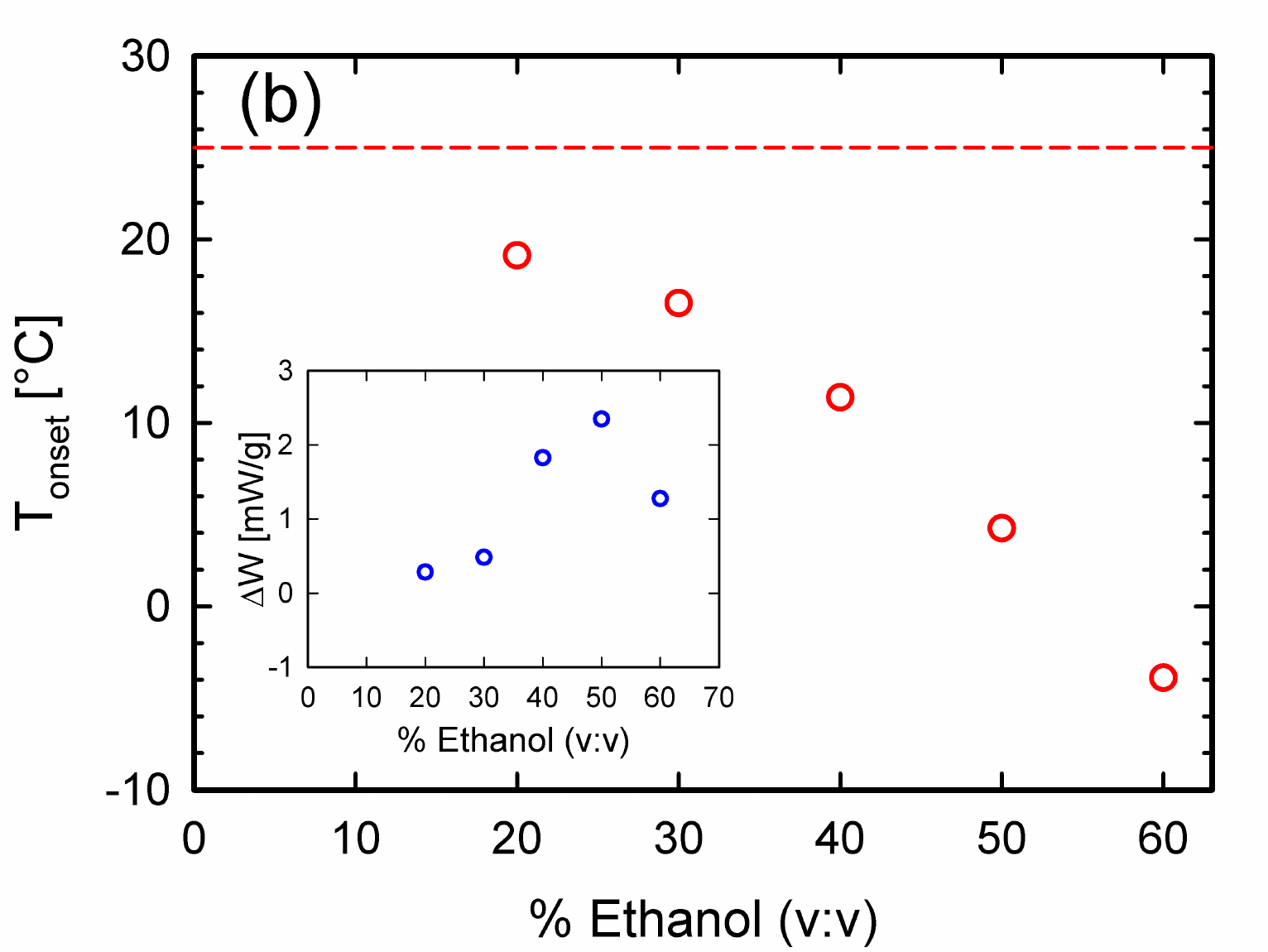}
\caption{\label{fig6} (a) Heat flow as a function of temperature for a sample prepared with a solvent comprising $50$ \% v:v  ethanol. The inset shows the phase transition of a gel prepared in a deuterated solvent with $30$\% v:v ethanol, from a transparent gel (right) to a turbid gel (left). The dashed gray line marks the onset temperature evaluated as shown in the main plot. (b) Phase transition temperature (main plot) and heat flow jump (inset) as a function of the solvent composition. The dashed horizontal line in the main plot indicates the temperature at which rheology experiments are performed.}
\end{figure}

All samples are transparent and look homogeneous to the naked eye at room temperature, but become white at low temperature (Figure~\ref{fig6}(a)).  The emergence of turbidity signs a liquid-liquid phase separation, which, however, does not lead to macroscopic phase separation due to sample viscoelasticity. The gluten proteins dispersed in ethanol-water binary mixtures exhibit thus an upper critical solution temperature behavior, as previously investigated by us in a study limited to a $50$ \% ethanol v:v solvent~\cite{banc_phase_2019}. \\
We use modulated differential scanning calorimetry (MDSC) to determine the phase separation temperature~\cite{arnauts1993,van_durme_influence_2004,seuring_polymers_2012,pincemaille_methods_2018}.  Figure \ref{fig6}(a) shows the heat flow as a function of temperature for a sample prepared with a solvent comprising $50$ \% v:v ethanol. A non-monotonic evolution of the heat flow is measured both upon cooling and upon heating, which is attributed to a liquid-liquid phase separation. The very low hysteresis (smaller than $1$ \textdegree C)  indicates a good reversibility of the transition. In the following, we focus on the cooling step, for which measurements start with the samples in the one-phase region. During cooling, the heat flow is first measured to gently decrease as the temperature $T$ decreases down to a temperature $T_{\text{onset}}$, at which the heat flow displays a jump of amplitude $\Delta W$ over about $3$ \textdegree C, before decreasing again smoothly as $T$  further decreases. $T_{\text{onset}}$ is always smaller than $25$\textdegree C, the temperature at which rheology measurements are carried out, indicating that visco-elasticity measurements are always performed on one-phase samples. We measure also that $T_{\text{onset}}$ decreases as the ethanol content of the solvent increases, from  $19.1$\textdegree C for a solvent with $20$ \% v:v ethanol  down to $-3.9$ \textdegree C for a solvent with $60$ v:v \% ethanol (Figure~\ref{fig6}(b)), in agreement with early observations for gluten proteins~\cite{dill_preparation_1925}. Hence binary solvents get better with increasing amounts of ethanol. They are better also than pure water, and than pure ethanol. For the latter, no homogeneous gel can be produced. These findings demonstrate thus a co-solvency effect. We also observe that the heat flow step decreases as the solvent becomes more depleted in ethanol (inset of Figure~\ref{fig6}(b)), apart from the sample with $60$ \% v:v ethanol. Consistently, for samples prepared with pure water and with a solvent comprising $10$ \% v:v ethanol, the enthalpic step is too small and the change in the heat flow cannot be accurately detected, thus preventing reliable measurements of a phase-separation temperature. Note that a vanishing $\Delta W$ might indicate that the liquid-liquid transition becomes second order as the solvent gets depleted in ethanol~\cite{hirotsu_critical_1988}.

\subsubsection{VSANS measurements}
Very small angle neutron scattering provides additional information regarding the impact of solvent composition on the gel structure.
For experiments performed with purely hydrogenated solvents, the main contrast probed by neutron scattering experiments is the one between the gluten chains and the solvent. The data obtained with pure water and with a solvent comprising $50$ \% v:v ethanol overlap. This indicates that, in the range of wave vectors probed experimentally ($3\times10^{-3}< q < 10^{-2}$ \AA$^{-1}$), hence corresponding to characteristic lengths $2\pi/q \approx 60-200$ nm, the gel structure is the same despite contrasting viscoelasticity, suggesting that these intermediate length scales and/or the contrast probed in this experiment are not the most relevant to understand viscoelasticity. Unfortunately, due to the weak scattering of hydrogenated samples, no reliable data could be obtained for $q < 3 \times 10^{-3}$ \AA$^{-1}$. However, previous data obtained with similar samples~\cite{JustineThesis} showed that the $q^{-2\pm0.2}$ power law evolution extends until $q=10^{-3}$ \AA$^{-1}$. In that case, the spatial organization of the proteins in the gels is characterized by a fractal dimension of $2$ at the supramolecular scale.

\begin{figure}[htbp]
\centering
\includegraphics[width=0.48\textwidth]{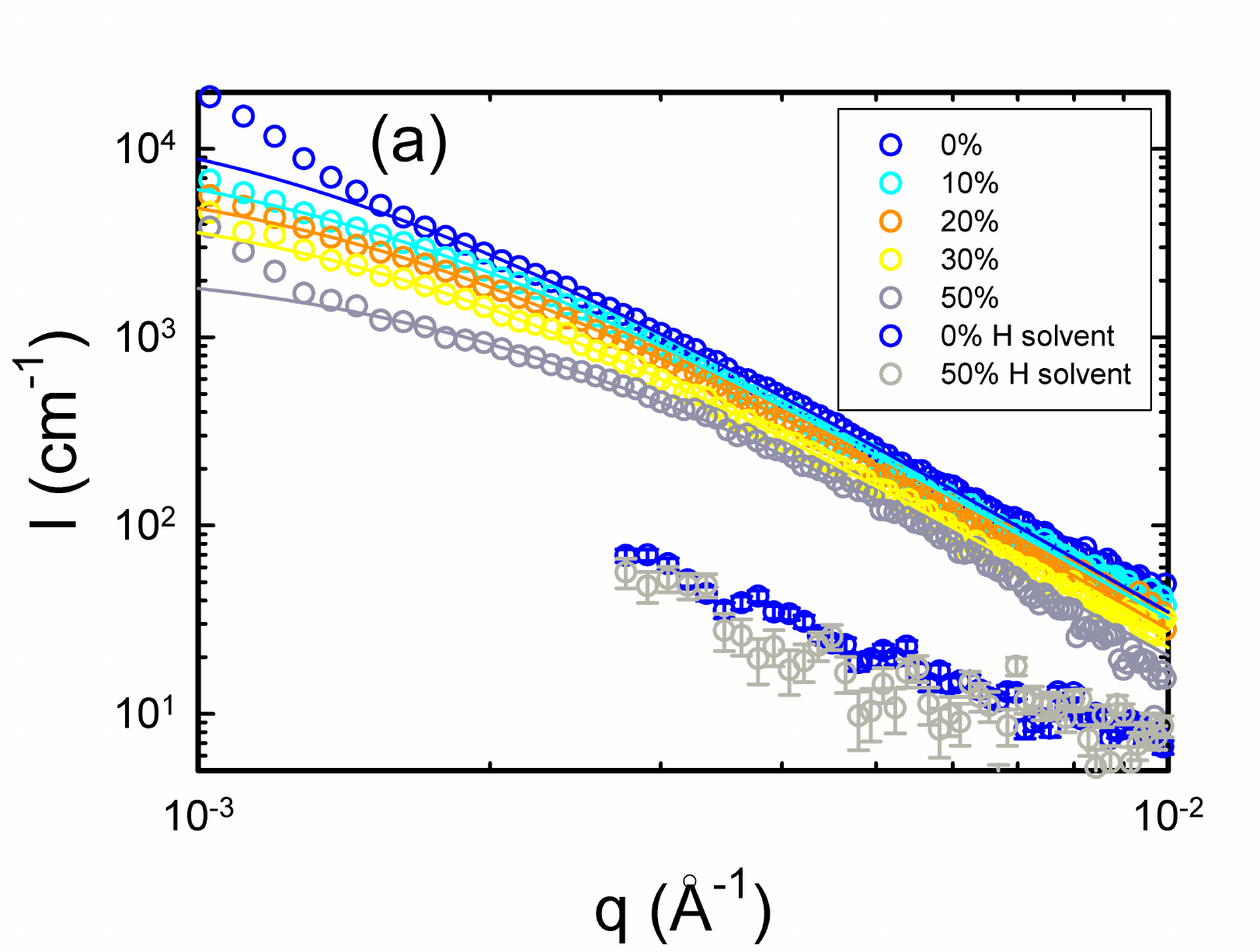}
\includegraphics[width=0.48\textwidth]{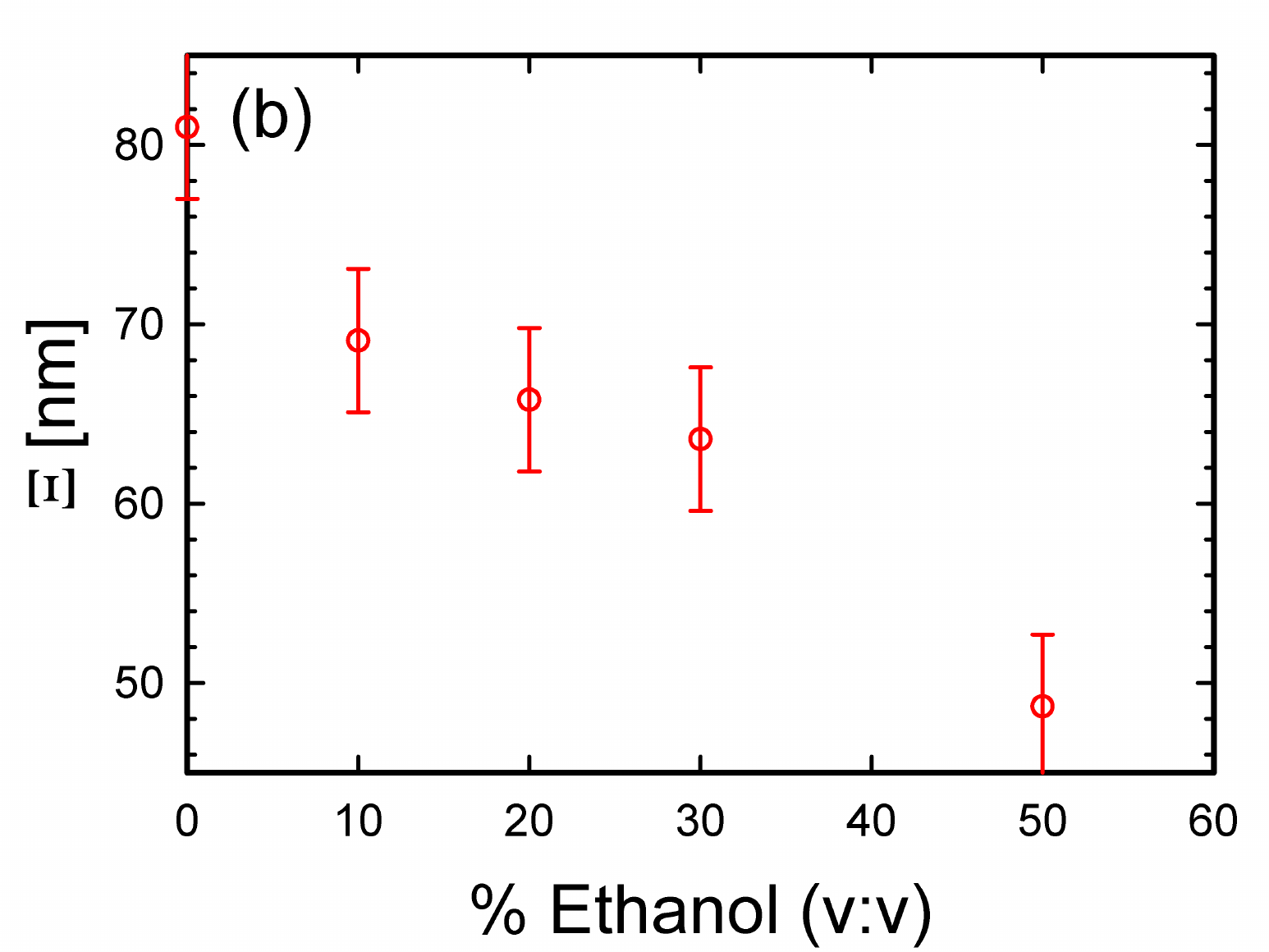}
\caption{\label{fig7}(a) Scattering intensity as a function of wave-vector for different samples as indicated in the legend. Symbols are data points and lines are fits (b) Variation of the characteristic length for deuteration inhomogeneities as a function of the solvent composition.}
\end{figure}

In order to evaluate the evolution of solvent quality with the ethanol content, we investigate gluten samples prepared with deuterated solvents. This approach is based on our previous results~\cite{banc2016}, where we showed that gluten samples prepared with deuterated $50$ \% v:v ethanol solvent display deuteration inhomogeneties on length scale of a few tens of nanometers, thus much larger  that any molecular length scales, while having the same fractal organization of the chains as hydrogenated samples (as inferred from small-angle X-ray scattering). We have attributed the deuteration inhomogeneties to regions rich in non-exchangeable hydrogen bonds because of H-bonding between proteins, thus preventing the natural H/D exchange between deuterated solvent and hydrogenated proteins~\cite{banc2016}. Considering that the isotope exchange is impacted by the solvent accessibility in proteins, the solvent quality could, in principle, be assessed through the investigation of deuteration heterogeneities. Figure~\ref{fig7}(a) shows that samples prepared with deuterated solvents display scattering patterns different from those prepared with hydrogenated solvent (as already observed by us for gels with smaller protein concentration and a $50$ \% v:v ethanol solvent). Scattering profiles are characterized by a power law decrease at large $q$ (with a power law exponent of the order of $-3$, always larger than the one measured for a hydrogenated solvent) and a cross-over toward a plateau at smaller $q$. The scattering profiles can be satisfactorily fitted by a generalized Debye Bueche model to extract a correlation length $\Xi$,  which can be regarded as the characteristic size of the regions rich in tight non exchangeable H-bonds: $I(q)=\frac{A}{\left[1+(q\Xi)^2\right]^p}$ with $A$ a prefactor to account for the contrast and $p=1.5$.

Figure~\ref{fig7}(b) displays the evolution of the correlation length, $\Xi$, with the ethanol content of the solvent. A clear decrease of $\Xi$ is measured, from $81$ nm for the sample prepared in pure water down to $49$ nm for the one prepared with ethanol $50$\% v:v. The decrease of this length with the addition of ethanol demonstrates a better H/D exchange, hence a better solvent accessibility to proteins.

\subsection{Discussion}

We investigate physical polymer gels produced by dispersing gluten proteins in water:ethanol binary mixtures, with varying ethanol contents, from pure water to $60$ \% v:v ethanol. The relevant role of hydrogen bonding in the formation of gluten gels is corroborated by the experimental evidence that urea addition induces a gel to sol transition~\cite{ng_power_2007,dahesh_spontaneous_2016}.
By calorimetry, we observe that the gels exhibit an upper critical solution temperature, and demonstrate a co-solvency effect. In particular, the solvent becomes better as it gets enriched in ethanol. Consequently, the upper critical solution temperature decreases as the amount of ethanol in the solvent increases. These findings are consistent with very small-angle neutron scattering (VSANS) data that show a better access of the solvent to the protein chains as the amount of ethanol increases.
The co-solvency effect of water:ethanol binary mixtures affects the strength of the gels at different solvent compositions. We find a time/solvent composition superposition for the linear viscoelasticity, allowing the building of unique master curves for the viscoelastic moduli. Such superposition principle is reminiscent of the time/cure superposition established for chemically~\cite{adolf1990} and physically crosslinking biological networks~\cite{richtering1992}. The master curves  for the loss and storage moduli present the hallmarks of the theoretical expectation for near-critical gel model above percolation~\cite{martin_viscoelasticity_1988,martin_viscoelasticity_1989}.
Self-similarity in the viscoelasticity implies a self-similar structure of the stress-bearing network, which is characterized by a single length scale, $\ell_c$. In our case, this implies, inter alia, a same fractal dimension for the network, independent of the solvent quality (Figure~\ref{fig7}(a)), in agreement with simulations~\cite{pandey_simulations_nodate}.
In principle, the solvation of proteins by water/ethanol could promote conformational changes\cite{amin_effect_2016} and possibly denaturation\cite{lamb}.
Here, however, no significant modification of amide bands in the infrared spectra is detected as the solvent composition varies (data not shown), implying no important changes of secondary structures, as opposed to other proteins (see for instance~\cite{renard_structural_1999} for a globular protein from milk where ethanol is found to induce protein denaturation).

\begin{figure}[htbp]
\centering
\includegraphics[width=1\textwidth]{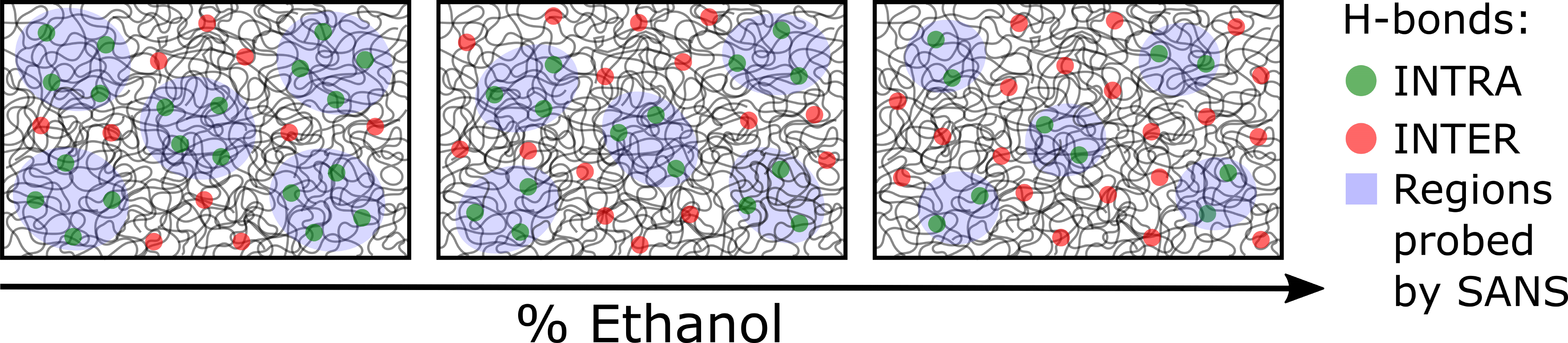}
\caption{\label{fig8} Schematic of the interplay between intramolecular and intermolecular bonding as a function of the ethanol content in the solvent. The red and green dots (not at scale) schematize the expected transition from intra- to intermolecular H-bonds. Intramolecular H-bonds are expected to be concentrated in the blue zones, whose size (measured by VSANS) decreases when the ethanol content increases.The background consists of a semi-dilute polymer solution, with a mesh size of the order of a few nm~\cite{dahesh2014}}.
\end{figure}

To rationalize our experimental findings, we propose the following physical picture, as schematized in the cartoon shown in Figure~\ref{fig8}. Because of the complex protein composition and the multiplicity of interactions at play, one can consider the sample as heterogeneous at a mesoscopic length scale, resulting from the interplay between two main types of interactions: H-bonds and hydrophobic interactions. One could thus expect a percolated network of well-solubilized polymer chains that are crosslinked by H-bonds in a background consisting of pieces of chains interacting  with hydrophobic interactions. Hydrophobic interactions would lock hydrogen bonds and prevent the building of interchain links, thus resulting in zones rich in intramolecular H-bonds. These zones would be the ones probed by VSANS. Ethanol molecules are less polar than water and thus weaken hydrophobic interactions  and promote polar interactions~\cite{castronuovo, castronuovo2}.
This would in turn facilitate the transition from intrachain associations to interchain hydrogen bonding, which governs the gel viscoelasticity.
Hydrophobic interactions are very sensitive to solvent composition. Conversely, the interactions (H-bonds) between well solvated chains, which confer elasticity to the network, are presumably not very sensitive to solvent composition.
At low ethanol concentration, hydrophobic intramolecular interactions might prevail over intermolecular hydrogen bonding. As a result, we would have a weak network arising from few intermolecular bonds spaced apart from each other (see Figure 8, left panel). As the ethanol content increases, solvation prevails over hydrophobic interactions. More chains would be  brought into the well-solvated phase and would  form more intermolecular bonding. The bonding density would increase and, subsequently, the characteristic length would decrease. As mentioned above, the decrease of the characteristic length is associated with a macroscopic increase of the elastic modulus. At length scales lower than the characteristic length of the network, self-similar dynamics are observed, as for systems in the vicinity of the sol-gel transition. Note that our experimental findings differ drastically from measurements performed on more simple gels (i.e. polyacrylamide gels) which comprises uniquely one type of chains. In this case, as the solvent quality becomes less good, the crosslinked chains collapse, yielding a collapse of the gel and solvent expulsion\cite{Tanaka1979}.
We finally mention that, in our experiment, gelation cannot result from microphase separation processes, as it might occurred for other types of polymer systems, because such processes implies a $q^{-4}$ scaling of the scattered intensity by the chains, at odds with our findings. \\
It is worth noting that the general concept of co-solvency effect on protein stabilization or denaturation is very complex and protein-specific as addition of a co-solvent might strengthen protein-protein interactions for some proteins  but weaken protein-protein interactions for others as it would do for synthetic polymers~\cite{spinozzi_proteins_2016}. Here we investigate complex protein mixtures comprising intrinsically disordered domains, which display many analogies with polymeric species and comprise also groups with different polarity and hydrophobicity. Co-solvency  can be viewed as a consequence of the polymer chains maintaining the optimum solvent environment, which will ensure maximum compatibility among the species (see for instance the review~\cite{zhang_polymers_2015} and references therein). This results in a peculiar interplay of different types of interactions (mainly hydrogen bonding and hydrophobic) in determining the rheology and self-assembly of this unique class of gels.

\section{Conclusions}

We have provided a rationalization of the interplay between co-solvency and gelation for a polymer gel made of gluten proteins.
Rheological tests show that the viscoelasticity of gluten gels at fixed concentration but different solvent composition can be put into context of the sol-gel transition of physical bonding systems. In such scenario, the rheological properties obey a time-solvent quality superposition principle that allows one to build robust post-gel master curves of the viscoelastic moduli, and to extract the critical gel exponent, as well as the equilibrium elastic modulus and the characteristic frequency of the gel, as a function of the water/ethanol content. The rationalization of the rheological response within the framework of the sol-gel transition suggests self-similarity of the gel network and an elastic modulus and a characteristic frequency which are related one to another. We find that, as the ethanol/water ratio increases, the self-similar gluten networks move further and further away from their gel point. The increase of the ethanol/water ratio increases the quality of the solvent for gluten gels, as demonstrated by differential scanning calorimetry and very small-angle neutron scattering (VSANS). In particular, the temperature for liquid-liquid phase separation decreases with increasing ethanol in the solvent, indicating better solvation, as confirmed by the VSANS scattering spectra of the different samples. The quality of the solvent directly affects the gelation dynamics of this class of systems, dictating the interplay between intramolecular and intermolecular interactions. At low ethanol content, intramolecular interactions prevail, resulting in a weak gel with large characteristic length. As the ethanol concentration increases, gluten molecules are better solvated and intermolecular hydrogen bonding is promoted, leading to stronger gels.\\
The outcome of this work is two fold. On the one hand, at a molecular level, this study provides further insight in the understanding of the co-solvency in determining gelation of complex polymeric systems such as gluten proteins. On the other hand, from an applications standpoint, this work provides the background to tailor the viscoelastic behaviour of polymeric gels in a food-grade solvent by adjusting the solvent quality.

\begin{acknowledgement}

Financial supports from ANR Elastobio (ANR 18 CE06 0012 01) and from  SAS PIVERT,
are acknowledged.  This work is also based upon experiments performed at the KWS-3
instrument operated by JCNS at the Heinz Maier-Leibnitz Zentrum
(MLZ), Garching, Germany.
The authors thank Joëlle Bonicel for HPLC measurements and Philippe Dieudonné for preliminary small-angle X-ray scattering measurements.

\end{acknowledgement}





\end{document}